# Content and Structure of Laboratory Packages for Software Engineering Experiments

Martín Solari [a, *]

Sira Vegas [b]

Natalia Juristo [b]

[a] Universidad ORT Uruguay, Cuareim 1451, 11100 Montevideo, Uruguay.
[b] Universidad Politécnica de Madrid, Campus de Montegancedo s/n, 28660 Madrid, Spain.
Email: martin.solari@ort.edu.uy (M. Solari), svegas@fi.upm.es (S. Vegas), natalia@fi.upm.es (N. Juristo)
[*] corresponding author

**Context:** Experiment replications play a central role in the scientific method. Although software engineering experimentation has matured a great deal, the number of experiment replications is still relatively small. Software engineering experiments are composed of complex concepts, procedures and artefacts. Laboratory packages are a means of transferring knowledge among researchers to facilitate experiment replications.

**Objective:** This paper investigates the experiment replication process to find out what information is needed to successfully replicate an experiment. Our objective is to propose the content and structure of laboratory packages for software engineering experiments.

**Method:** We evaluated seven replications of three different families of experiments. Each replication had a different experimenter who was, at the time, unfamiliar with the experiment. During the first iterations of the study, we identified experimental incidents and then proposed a laboratory package structure that addressed these incidents, including document usability improvements. We used the later iterations to validate and generalize the laboratory package structure for use in all software engineering experiments. We aimed to solve a specific problem, while at the same time looking at how to contribute to the body of knowledge on laboratory packages.

**Results:** We generated a laboratory package for three different experiments. These packages eased the replication of the respective experiments. The evaluation that we conducted shows that the laboratory package proposal is acceptable and reduces the effort currently required to replicate experiments in software engineering.

**Conclusion:** We think that the content and structure that we propose for laboratory packages can be useful for other software engineering experiments.

# 1 Introduction

Experimentation is a way of maturing knowledge. It is at the heart of the scientific method. Experiments manipulate and observe reality in a controlled, rigorous and systematic manner in order to discover variables that somehow influence a specified phenomenon [1]. For this approach to be successful, experimentation must be iterative, taking earlier results to refine procedures for studying the phenomenon in more depth. As a research method in software engineering (SE), experimentation dates back only a few decades having first been formalized around 1986 [2].

Replication is a way of exploring experiments in more breadth and depth [3]. Replications serve two purposes: (i) they may study whether the results are valid for other contexts or whether other researchers are able to get the same results, and (ii) they may, if changes are made to an experiment, allow studying influencing variables and provide information about the bounds of the acquired knowledge [4]. Since replication is troublesome in SE, some researchers [5] have proposed alternative methods. Although there are many methodological uncertainties surrounding replication, its importance is clear [6].

Experiments are composed of concepts and of particular artefacts for experiment operation. The conceptual elements include the constructs, hypotheses and experimental design. The operational elements include the instruments, materials and procedures. On this ground, an experiment is a highly complex entity both conceptually and operationally. This is an obstacle to its replication, particularly when a researcher is replicating an experiment that he or she did not design [7].

There are some ways of facilitating experiment replication. They include collaboration among researchers and laboratory packages [8] [9] [10]. The use of such types of knowledge transfer mechanisms is widespread among researchers from other disciplines, such as physics, biology and the social sciences [11] [12] [13].

Laboratory packages (LP) contain the information and materials required to replicate an experiment. The term protocol is often used in natural science [14] to refer to the packaging of key information about an experiment for its transmission and use in a replication by others. Whereas the term protocol underscores strict adherence to a procedure, the term LP focuses on the transfer of information in self-contained units. Both components are necessary for conducting a replication.

The content of a LP is not static; it needs to be adapted to the needs of the researcher running the replication. A LP is usually task driven (that is, written like a *recipe*) and may use alternatives other than text like video or software. A scientific paper is unlikely to include this type of content as there are usually constraints on space and format [15]. Scientific publications usually stress the high-level conceptual aspects, omitting more specific operational issues that are helpful for replicating experiments. In some disciplines, this gap has been filled by journals dealing specifically with experimental protocols like *Nature Protocols*[1]. Open digital repositories of experimental protocols have also been set up, providing for discussion among researchers based on a forum system[2]. These repositories enable a faster information exchange than journals, as well as the use of alternative formats for answering questions and helping other researchers to replicate the experiment.

---

[1] http://www.nature.com/nprot/index.html

Although SE has made significant progress with regard to the conceptualization and running of replications, information packaging and transfer are still an open question. The use of LPs has been defended ever since the earliest systematic experiences of SE experiment replication [7]. The need for LPs as a key component for conducting families of experiments was again stated later on [8]. Collaboration between groups of SE experimenters was the focus of research project that proposed the use of a knowledge transfer model also based on LPs [16]. Interest in the topic appears to have cooled within the SE community, but this does not mean that how to package and transfer the information needed to run a replication has been solved. Although the number of replications has grown over the last few years, it is still not a common practice, even less so if researchers are trying repeat an experiment at other sites [17]. Neither are there many cases of SE experiments whose materials are published for use by other researchers, and experiments whose procedures are shared with a view to conducting exact replications are even rarer still [18].

We aim to identify which information a LP for SE experiments should contain. Proper LPs would encourage more and better replications. Replications not only contribute to the body of SE knowledge, but also help researchers to learn about how to experiment with and observe the phenomenon. This helps to mature SE experimentation.

Our immediate goal is to improve the LPs used to replicate experiments within a SE experiment network. Although the research was undertaken in a small experimental research network, we believe this context is representative of the current replication and information exchange scenario between experimental researchers in SE.

We studied three experiments and seven replications iteratively. First, we observed the incidents in replications of one experiment run by researchers belonging to several research groups. From these observations we derived a preliminary LP structure. In later iterations, we refined the structure by observing more replications of the same experiment run by researchers belonging to the same and other research groups. Finally, we generalized the structure to build LPs for other two experiments.

The structure of the paper is as follows. Section 2 discusses related work, analysing the use of LPs and experimental protocols first in other disciplines and then in SE. Section 3 describes the research method. The other sections each address one of the stages of the research. Section 4 reports the diagnosis based on the observed incidents in experiment replications. Section 5 outlines the proposed LP structure. Section 6 describes the LPs generation for three experiments. Section 7 describes how LPs were evaluated by observing replications using them. Section 8 reflects on how the proposed LP might be generalized. Section 9 present further evolution plans for our proposal. Finally, Section 10 outlines the conclusions of the research.

---

[2] http://www.nature.com/protocolexchange/

# 2 Related Work

## 2.1 LPs in other disciplines

It is common practice in many scientific disciplines to use LPs in different types of experimental replications. In the social sciences, for example, it is usual to share sampling, observation or intervention procedures for complex phenomena [13]. Disciplines like biology and medicine use repositories containing the protocols to be followed by experimenters [19]. These protocols facilitate the aggregation of studies replicated by different researchers at different sites.

Since the 1990s several scientific communities have helped to create new methods for speeding up research and promoting information exchange [20]. Digital media are not subject to the typical constraints on number of pages and format imposed by the print media. They also facilitate discussion and contacts among researchers.

Giles [15] argued in an opinion column that researchers visit each other to learn the protocol because much of the knowledge about the experiment is tacit. The column cites several scientific publication editors who stress the need to publish more detailed information and provide for direct exchange among the researchers that are going to run replications. Examples of journals that provide these exchange mechanisms are: *PLoS ONE*[3], *Cold Spring Harbor Protocols*[4] and *Nature Protocols*[5].

Protocol publication and study preregistration have become the norm in medicine. In recognition of the importance of detail in order to ensure the rigour of clinical trials, several initiatives try to standardize the content of such reports and protocols. The CONSORT (CONsolidated Standards of Reporting Trials) initiative promotes study reporting quality by means of a content checklist that was originally proposed in 1997 and has been updated on several occasions [21].

There is agreement among the editors of several medical journals about the importance of having access to the full protocol of published trials [14] [22]. Journals like the *Lancet*[6], *British Medical Journal*[7] and *PLoS Medicine*[8] demand that the protocols be submitted together with controlled experiment reports for inclusion in the review process. However, current guidelines on protocol content vary substantially with respect to recommendations [23]. Recently, the SPIRIT (Standard Protocol Items: Recommendations for Interventional Trials) initiative proposed the standardization of the protocol content of experiments in medicine [24].

---

[3] http://www.plosone.org/

[4] http://cshprotocols.cshlp.org/

[5] http://www.nature.com/nprot/index.html

[6] http://www.thelancet.com/protocol-reviews

[7] http://www.bmj.com/about-bmj/resources-authors

[8] http://www.plosmedicine.org/static/guidelines

An exceptional example of LP or protocol repository is *Nature Protocol Exchange*[9]. This repository provides for the open exchange of experimental protocols and an open discussion. Replicating experimenters can submit queries and baseline experiment authors can provide the details that they had omitted to specify. Unlike *Nature Protocols*, the repository does not follow a process of arbitration before the protocols are uploaded. Research groups can sign up to share their own protocols or query the protocols of other groups. The repository is open to experiments from any scientific discipline. These procedures and data can be taken by other researchers to run replications or to perform secondary studies (systematic literature review (SLR), meta-analysis, etc.).

Although there is no mandatory structure, the suggested format for uploading protocols is the same as for submissions to *Nature Protocols*. This protocol format is composed of the standard sections of a scientific paper (authors, abstract, introduction, etc.), but also by sections specializing in experiment replication like:

- Procedure
- Timing
- Troubleshooting
- Anticipated Results
- Supplementary Information

Step-by-step guidance should be provided for the Procedure section. The Supplementary Information section may include almost any type of digital information like audio, video, equations, methods, figures, etc. This type of content clearly goes beyond the dissemination of the findings, aiming to record detailed procedures and materials for other experimenters to be able to replicate the experiment.

The *Nature Protocol Exchange* supports a more fluent discussion between researchers conducting the same experiment because it accommodates different types of materials, enabling and encouraging the inclusion of more detailed information about procedures and provides a rapid exchange process. Note that all these initiatives share the following implications:

- Detailed information should be supplied for an experiment to be able to be replicated,

- Information about an experiment is continuously evolving [15].

A LP follows the dynamics of the experimental process. Rather than a closed file, it is an open information exchange tool within a research process. Replications are a means for both building confidence in the results and learning new variables about the phenomenon under study [25]. Replications are also a way to learn about and refine the methods of observation, that is, learn more about the experiment itself. On this ground, the interaction between LPs and the researchers that use them is complex, and they are an intrinsic and evolving part of experimental research.

---

[9] http://www.nature.com/protocolexchange/

## 2.2 LPs in SE

The LP issue has been addressed on several occasions by the SE community. The community recognized the need for LPs to promote more replications in the 1990s [7] [26]. In the late 1990s, Basili proposed the concept of families of experiments as a way of achieving coordinated replications based on a LP [8].

Shull revived the idea of transferring knowledge among researchers when he published one of the most sophisticated examples of LP. This LP, published in web format, accounts for multiple replications run within the same framework. Apart from the LP itself, Shull et al. propose an explicit knowledge management model to facilitate information exchange within the replication process [27].

After the publication of this LP, the SE community's interest in this topic waned. Although these early papers succeeded in raising the awareness of many experimenters about the need to publish the experimental materials for the purposes of replication, the experiments run do not tend in practice to provide LPs that are as mature as the one proposed by Shull. Many published experiments are supplemented by information available on the web, but hardly any include anything more than the experimental material (for example, programs, data collection sheets) or raw data supporting the results [18]. Neither of these options is sufficient for running a replication.

Proposals for standardizing controlled experiment reporting were published in the early 21st century [28]. They were well received by the SE community. This proposed reporting format contains specific sections constituting the groundwork for a publishable paper. However, the focus on structuring the publication of experiment results omits the materials and procedures required for replication. A preliminary proposal specifically for reporting experiment replications was made later [29], but again it does not include a structure for publishing materials and procedures to enable replication to continue.

Recently a few groups have systematized the publication of materials and procedures for their experiments, making the information available on the web. Two notable examples are the Software Engineering Research Group's (SERG)[10] site at Lund University, which contains LPs for more than 20 experiments, and Lutz Prechelt's page at the Free University of Berlin, which contains a section with materials and data for replicating eight experiments[11]. The structure of the LPs published at these sites is ad hoc and depends on the experiment. LPs are not standardized even within the same group. Both groups have encouraged the replication of their experiments and put the accent on information sharing with other researchers.

The systematic mapping study conducted by Da Silva reported 133 replications of SE experiments. They found that the number of replications is growing, and there is more and more discussion about the importance of having a collaborative agenda among researchers [17]. However, the mapping does not identify any theoretical inputs for building LPs, except for the ones mentioned at the beginning of this section. They state that many examples of LPs have been published, but problems such as the proper recognition of the contributions of different researchers persist [30].

---

[10] http://serg.cs.lth.se/index.php?id=32563

[11] http://page.mi.fu-berlin.de/prechelt/Biblio/#package

In order to form a closer idea of the state of LP practice with regard to SE experiments, we have extended the review that we published back in 2006 [18]. We searched the public web (using the Google search engine) instead of confining the results to scientific bibliographic databases. The keywords used for the search were: *software engineering replication* (alternative *laboratory*) *package*. We manually reviewed the primary results to see which really did contain LPs. We also added LPs with which we were personally acquainted about to the review manually. We only considered LPs that were publicly available online (which is the only way of guaranteeing access for potential replicating experimenters). Some of the LPs were not available at the time that we concluded this research. However, they are considered because they were available when we conducted the original search in 2006 [18], updated in 2010 [31] and again in 2014 for this paper.

We divide the identified SE experiment LPs into three categories: single experiments, material repositories and general-purpose support infrastructures:

- LPs built for a **single experiment** or single family of experiments usually have a specialized structure suitable for describing the particular characteristics of the experiment. It is common practice for this type of LP to be referenced from scientific papers that, for reasons of space, cannot include all the details of the experiment. An example of this type of LP is the UML comprehension experiment published by Genero et al. [32].

- **Material repositories** are a special case of LP focusing on experimental objects. For example, they contain different versions of programs and their respective information. These objects can be used in different types of experiments. An example of this type of LP is the repository of artefacts for testing experimentation created by Do et al. [33].

- **General-purpose support infrastructures** are not confined to a specific family of experiments. On the contrary, they can support the activities of the experimental process for any family of experiments. The structures may be conceptual (for example, data models, ontologies) or be implemented in a specified system. We have only considered support repositories or infrastructures that have at least one instance of a specific experiment. An example of this type of LP is the computational framework proposed by Mian et al. [34].

Our research will focus on the first type of LP. We believe that a LP should be a self-contained tool that is independent of other documents. On this ground, we examine the information that the LP contains rather than information that is scattered across articles published about different experiments. This does not mean that other types of LPs are not needed in order to take replication forward, but merely that they are beyond the scope of our research at this stage.

Table 1 gives an overview of the LPs identified for single experiments. We found that the number of publicly available LPs is rather low compared with the number of experiments and replications in SE [35] [17]. Whereas material or additional descriptions may have been published in some medium or another, they are not easy for web search engines to locate. The repositories of SERG at Lund University and Prechelt's page at Fee University of Berlin were not found in our initial search and comparison. Both repositories have LPs for multiple experiments and significantly improve the state of published material in SE. However, the repositories don't use uniform LP structures.

**Table 1: Overview of identified LPs.**

| LP | 1 | 2 | 3 | 4 | 5 | 6 | 7 |
|---|---|---|---|---|---|---|---|
| *Authors* | Kamsties & Lott [36] | Basili et al. [37] Shull et al. [16] | Thelin et al. [38] | Dunsmore [39] | Do et al. [33] | Du Bois [40] | Genero et al. [32] |
| *Year* | 1995 | 1995 2004 | 2003 | 2003 | 2005 | 2006 | 2008 |
| *Research topic* | Functional, and structural testing, code review | Perspective-based reading | Checklist-based reading | Object-oriented inspection | Regression testing | Class decomposition | UML stereotypes |
| *Known replications (institution)* | Kaiserslautern Strathclyde | Maryland NASA SEL Kaiserslautern | Lund | Strathclyde | Nebraska | Antwerp | Bari Castile-La Mancha |

Most of the identified LPs are appendices to published papers. On one hand, scientific papers must adhere to journal or conference formats and space constraints and often do not include specific replication guidelines.

We found that the LPs also include scant descriptions. In most of the cases, these descriptions do not target the replicating experimenter but are designed to provide materials or data to supplement a paper. While this is of no assistance for encouraging replication, it does help to make experimental results more transparent. The sections of the analysed LPs differ substantially and do not account for different experimental process activities in the same manner. We have also found differences in the language used to describe the same experimental concepts.

An all-round analysis of the identified LPs gives an idea of the key weaknesses of SE LPs:

- Most LPs focus on the artefacts for operating the experiments. Some do not contain any more information than the operational material. Generally, this is because they assume that experimenters will have read a paper before consulting the LP. In these cases, we suggest that the term material be used to make it clear that it has a different objective than LPs. In fact, such material might be referred to as supplementary information to make its purpose clear.

- Descriptions and guidelines for replication activities are less common. Instructions targeting replicating experimenters have only ever once been built into the LP format itself [16].

- No LPs have included information on replications conducted in the past or the evolution of the experimental research (extracted pieces of knowledge and their interpretation), save on the occasion of one experiment [16]. The replication findings are usually published as part of papers or technical reports but are not incorporated with all the other information into one and the same structure.
- LPs do not have a standard structure, and different formats are used to organize the content. Although most of the studied LPs are based on a web page and links to files in other formats, there are a vast number of formats and configurations.

Therefore, the problem of how to package the information required to replicate an experiment in SE can by no means be considered to have been solved. Yet we build our solution on some of the previous research.

The starting point of our proposal is the analysis of previously published LPs for SE experiments. Most of these LPs include the experiment materials as an appendix to a published article. We found that scientific journals had space and format limitations, and the materials did not include any replication guidelines. Therefore, we wanted to address the problem using a self-contained structure.

One notable exception is Shull et al.'s LP [16], which is specifically constructed to help different groups replicate an experiment. This LP provides a description of the experiment as separate from the report for each replication. It also includes a history section reporting how the experiment evolved. This LP had a considerable influence in our proposal. The structure that we propose reuses three key components already present in Shull et al.'s LP:

- includes specific guidelines for replicating the experiment,
- separates the description of the baseline experiment from the report for each replication,
- summarizes the evolution of the experiment inside the LP structure.

## 3   Research method

The need of LP emerged as a necessity when internally and externally replicating an experiment. Over 14 years ago, the experimental research group at the Technical University of Madrid (UPM)[12] contributed to a family of testing experiments [41]. We received the materials from one of their replications [42] which was based on a previously published LP [36]. In the first place, we had to study and organize these materials to produce an experimental design adapted to the particular conditions of the UPM replication. This experiment was internally replicated three times by the same researchers [43]. Later, this experiment was again internally replicated (at UPM) but by other researchers from the same group. This meant that a LP had to be put together to transfer the information. At a later stage, the experiment was replicated under an arrangement with other groups and again the LP had to be adapted in order to conduct external replications [9] at Universidad de Sevilla (UdS), Universitat Politècnica de València (UPV) and Universidad ORT Uruguay (ORT). The experiment was modified according to the recorded outcomes, and the LP had to include this new information [44]. Other groups are still replicating the experiment using the LP [45] [46] [47].

---

[12] http://www.grise.upm.es/

This research commenced in 2006 with the definition of a framework for observing experimental replications and evaluating LPs [48]. We applied this framework in each iteration of the research. Thanks to the evaluation framework, we were able to identify any incidents occurring that were an obstacle to or jeopardized the results of the replication. We thoroughly examined each particular incident that occurred, relating the incidents across different replications. The incidents were grouped for later generalization in order to identify archetypal incidents likely to occur when replicating experiments in SE.

The major sources of information were the researchers participating in the replication of the experiment and the LPs used in each replication (including documents and any form of communication). The researchers were interviewed after a replication. The strategy underlying the research iterations is to gradually alter experiments and replication sites. Table 2 shows the iterations in the research method and the different experiments and research groups involved in each iteration.

**Table 2. Overview of the research iterations.**

|  | **Iteration 1** | **Iteration 2** | **Iteration 3** | **Iteration 4** |
|---|---|---|---|---|
| **Experiment** | *DefDetect* | *DefDetect* | *DefDetect* | *UnitTest and TeamPersInf* |
| **Replications** | *UPM 2000* | *UPM 2004, UdS 2005, ORT 2005, UPV 2006* | *ORT 2009, UPM 2011* |  |
| **Diagnosis** | Experimental incident analysis | Experimental incident analysis | Experimental incident analysis |  |
| **Intervention** |  | Draft of proposed LP to help mitigate identified incidents | LP instantiation | LP instantiation |
| **Evaluation** |  |  | Evaluation of LP use in experiment replications | Evaluation of the instantiation of the proposed LP |

In Iteration 1, we ran a diagnosis of the replications of an experiment about defect detection techniques (DefDetect experiment) run within our research group [49]. We had replicated the experiment using a LP provided by another group [42]. After analysing what information we would have liked to have had about the experiment for replication, we generated a new LP for this same experiment including this information for use in successive replications.

In Iteration 2, we extended the diagnosis to replications of the same experiment conducted at other sites. The external replications [9] used an improved version of the LP. We devised a LP proposal that met the needs revealed by these three new replications. We used the identified experimental incidents to propose the structure of a LP to mitigate such incidents in replications generating a new LP version for the DefDetect experiment. In iteration 3, we evaluated the experimental incidents in replications using the new LP.

Once the structure and the LP structure had been stabilized for the DefDetect experiment, we moved on to other experiments with the aim of generalizing the proposed LP structure. In Iteration 4, we validated and refined our proposed LP, generating LPs for another two experiments run and replicated by other research groups. These experiments and research groups were progressively further distant from the original experiment and group. In particular, we generated LPs (based on the structure proposed in Iteration 2) for another two experiments: another experiment by the UPM group (conducted by different researchers than the DefDetect experimenters) on team and personality influence in development (TeamPersInf experiment) replicated at [50] and an experiment on unit testing techniques (UnitTest experiment) [51] by a research group that is not a member of the original network (University of the Republic, Uruguay).

# 4   Analysis of experimental incidents

In the diagnosis stage, we analysed experimental incidents in several replications. The analysis focused on the problems and obstacles occurring in each individual replication. Researchers running replications all have different backgrounds in terms of both experimentation and the experiment domain. Therefore, they have different information needs when it comes to replicating an experiment. On this ground, their subjective views of the process of replication have to be taken into account. As the phenomenon in question is complex, we used a qualitative evaluation method. Evaluation is empirical as we observe the use of the LP as part of several separate replications (one at a time).

During the evaluation we analysed the instruments and, particularly, the LP used during the replication. We identified and studied particular experimental incidents and tried to determine the extent to which the LP influences each incident [48]. We classified the incidents by types and identified the needs that the LP was expected to satisfy. The specific goals of the evaluation of the replications were:

1) Discover the purpose and setting in which the replication was run,

2) Discover the characteristics of the LP used, as well as other instruments of communication used to run the replication,

3) Identify the incidents in each experimental activity,

4) Understand the viewpoint of the replicating experimenter with respect to the LP and to the development of the replication.

In order not to interfere in the replication, the evaluation was conducted post-mortem at the end of each replication. As replication is a highly intellectual activity, mere observation (without interaction with experimenters) did not show up enough information in order to understand what was going on. The key information-gathering activity was a semi-structured face-to-face interview [52] with the experimenter responsible for each replication. Thanks to this interview method, we were able to gather a considerable amount of information from the key source and rapidly answer any questions. Before the interview, all available documents associated with the replication, including the replication LP itself, were collected. This preliminary phase improved the effectiveness of the interviews as it put things into context. Also, experimenters were contacted by email if any further questions arose at a later date. The full responses were used to build a systematic transcript and then a structured summary for each replication. The summary uses coded fields to describe each replication using a metalanguage that evens out the differences between the literal responses of each experimenter.

The coding was applied iteratively to output a list of experimental incidents. The codes of each incident were not generated a priori; they were output iteratively based on observations made during the evaluation of the replications. First, one researcher encoded the incidents of each particular replication and then the same researcher analysed the similarities with the other evaluated replications. The same code was used to denote identified incidents that are comparable to each other. Another researcher reviewed the mappings between the generated codes and the systematic transcripts. As a result, instead of the literal expressions used by the replicating experimenter during the interview, the detected incidents are designated using a standardized code that stands for related problems and obstacles.

We grouped the incidents by categories in accordance with the experimental process activities. Table 3 summarizes the identified incidents. Table 3 also shows the output of the diagnosis stage for the first two iterations. For a detailed description of each incident and the process enacted for their identification, see [48].

The LP and inter-researcher communication mechanisms used in each replication were different even in replications of the same experiment conducted close together in time. The standard LP for the networked replications from 2005 to 2006 consisted of a main document and related files. But this document evolved over time, and each version omitted some information. Some elements were communicated by email or at meetings depending on the replication. These variations led to specific incidents, such as the failure to include defect information for programs or some tasks requiring more effort. Therefore, the definition of incident is dependent on the context of each replication. What is considered to be an incident in some scenarios will not necessarily be an incident in others.

The identified incidents are always based on the viewpoint of the researcher who is replicating the experiment in a specific context. This means that experimenters with different backgrounds may identify different incidents in some contexts. For example, irrespective of the information included in the LP, a researcher could be unfamiliar with the experiment objects. For instance, none of the LP versions contained guidelines for data analysis. In a replication where researchers are already familiar with data analysis techniques or another group performed the data analysis, this may be not perceived as an incident, since the incident is defined from the researcher's perspective.

**Table 3: Incidents identified across replications.**

| Activity | Incidents | Iteration 1 | | Iteration 2 | | |
|---|---|---|---|---|---|---|
| | | UPM 2000 | UPM 2004 | UdS 2005 | ORT 2005 | UPV 2006 |
| ***Communication*** | *No communication* | X | | | | |
| | *Communication without direct contact* | | | X | | |
| | *Limited or after-the-event communication* | | | | X | |
| | *No meeting held to validate design* | | | X | X | |
| | *No in-person observation of sessions* | | | X | X | |
| | *Replicator felt cut off* | | | X | | |
| | *There were a lot of misgivings* | X | | | X | |
| | *Experiment is hard to understand* | | X | | | |
| ***Experimental design*** | *Less time assigned to one of the treatments* | | | | | X |
| | *Factor crossing error* | | | X | | X |
| | *Non-randomized assignment of subgroups* | | | X | X | X |
| | *Unbalanced experimental groups* | | | X | | |
| | *Misgivings about the impact of the design change* | X | | | X | |
| | *Underestimated experiment adaptation workload* | X | | | X | |
| ***Training*** | *Reduced training time* | | | X | X | X |
| | *Other material used in training* | X | | | X | X |
| | *Training received did not match the treatment* | | | X | | X |
| | *A treatment (functional technique) was misunderstood* | | | X | | |
| | *A treatment (structural technique) was misunderstood* | | | | X | X |
| ***Material preparation*** | *Too much material for one treatment* | | | | | X |
| | *The supplementary sheet was removed* | | | X | | |
| | *Line numbers were added to the source code* | | | | X | X |
| | *One of the programs could not be compiled on the platform* | X | | | | |
| | *Support from other people was required to prepare the material* | | X | | | X |
| | *Underestimated material preparation workload* | | | X | X | |
| | *Fear that material contingencies would affect session time* | | | | X | X |
| | *Complex, time-consuming activity* | X | | X | X | X |
| ***Operation*** | *Participants ask a lot of questions* | | | X | X | X |
| | *Participants do not read all the material* | | | | | X |
| | *Researcher is unfamiliar with objects and cannot answer questions* | | X | | | X |
| | *Limited session time* | | | X | X | |
| | *Rigorous atmosphere of session* | | | | X | |
| | *Effect of fatigue on participants* | | | | X | |
| | *Some participants apply wrong treatment* | X | | | | |
| ***Data analysis*** | *No analysis was conducted* | | | X | | |
| | *Analysis was postponed* | | | | X | |
| | *Support from another person was required to complete analysis* | | | | | X |
| | *Correctness criterion defined by experimenter* | | | X | X | X |
| | *Test cases and faults written by participants are ambiguous* | X | X | | | X |
| | *Correction was perceived as very complicated* | | X | | | X |
| | *No guidelines or examples were available for the analysis* | | | | | X |
| | *No previous data were available to compare results* | X | | | | |
| | *Fault description was not integrated into the laboratory package* | | | | | X |
| | *Fault description was not detailed enough* | X | | | | |
| | *Errors in the fault/failure description* | | | X | | |
| ***Research process*** | *Notes taken on the replication were destroyed* | | X | | X | X |
| | *Replication reporting was postponed* | X | | X | X | |
| | *Research cycle was incomplete* | | | X | X | X |
| | *Experiment was not evolved* | | | X | X | X |

# 5 Designing a LP proposal

The LP proposal is composed of a set of elements aiming to improve the information available to the replicating experimenter and avoiding incidents in the replication. The proposed LP supports *all* the activities of the experimentation process and not only execution. In order to identify the activities of the experimental process, we used the proposals in different experimentation books [53] [1]. In order to provide a complete description of the experiment, the design of our LP proposal considers Jedlitschka's guidelines for reporting experiments [28] as well as Carver's specific guidelines for reporting replications [29]. We also considered facets of the technical documentation quality [54] [55] [56]. These facets include overall usability of the document containing the LP, information navigability, search and retrievability. These cross-cutting issues are evaluated in each version of the instantiated LP. The complete description of the proposed LP is available as a web appendix in our website[13].

For the initial design of the LP proposal, we used the experimental incident list as main guidance for generating specific components in the structure. We analysed experimental incidents by category and hypothesized which components could have prevented each incident. After generating an initial design, we mapped each one of the identified incidents to one or more components of the proposed LP in a comprehensive relationship matrix. This matrix covered all identified experimental incidents to assure that each one was addressed by at least one component of the proposal. This relationship matrix is included in the web appendix.

## 5.1 Proposed LP structure

Our LP proposal is divided into modules. We think experimenters might have very diverse information needs. Each experimenter has a different background and distinct experience in experimentation. We have found that an experiment is usually run by several researchers who play different roles. The structure of a LP should enable each researcher to easily identify the key information for the task that he or she is performing without having to leaf through pages of information that have nothing to do with the responsibilities of his or her role. The aim of dividing the LP into modules is to cover all the phases of the experimental process and meet the information needs of different researchers depending on what activity of the experimentation process they are to perform in a manner that is clear to all LP users.

The modules constitute separate parts of the document. However, they are explicitly linked to each other so that anyone who wants to read all the documentation instead of focusing on a particular activity will find it easy to navigate from one to another. We see a module as a structure that is used separately for some experimental activity. For example, the replicating experimenters use one module to train subjects, another to adapt the design and another to record the results of the replication. We have tried to make each module as self-contained as possible in order to satisfy information needs without the researcher having to resort to other parts of the documentation. If necessary, the modules are further divided in order to include more specific information on diverse subtasks. For example, the experiment module is subdivided into sections on experimental design, session operation and data analysis.

---

[13] http://www.grise.upm.es/LaboratoryPackage

We expect that running a replication should also imply upgrading the LP to integrate the new piece of generated evidence that will help to mature the knowledge pursued by the family of experiments (and supported by the LP). After a replication, it should at least be added to the experiment log generating a new module containing the detailed description of the replication and its results. As a result of the replication, however, observations are often made and lessons learned about the experiment itself, the support materials or even the theory. Therefore, a replication may generate a new version of the LP modules containing this information. Note that the generation of new versions of modules should never result in the replacement of earlier versions, that is, the package also acts as a historical record.

Figure 1 shows the main structure of the LP with the name of each module. The directed lines denote the dependencies between modules. This entire structure is the starting point. Elements can then be added to the LP to mitigate the experimental incidents that we have identified. We have tried to design a preliminary structure that is as detailed as possible to cover the information needs and experimentation process activities. Each module is subdivided into sections. These sections provide help for mitigating the missing information detected in the incidents of a particular experimental subtask.

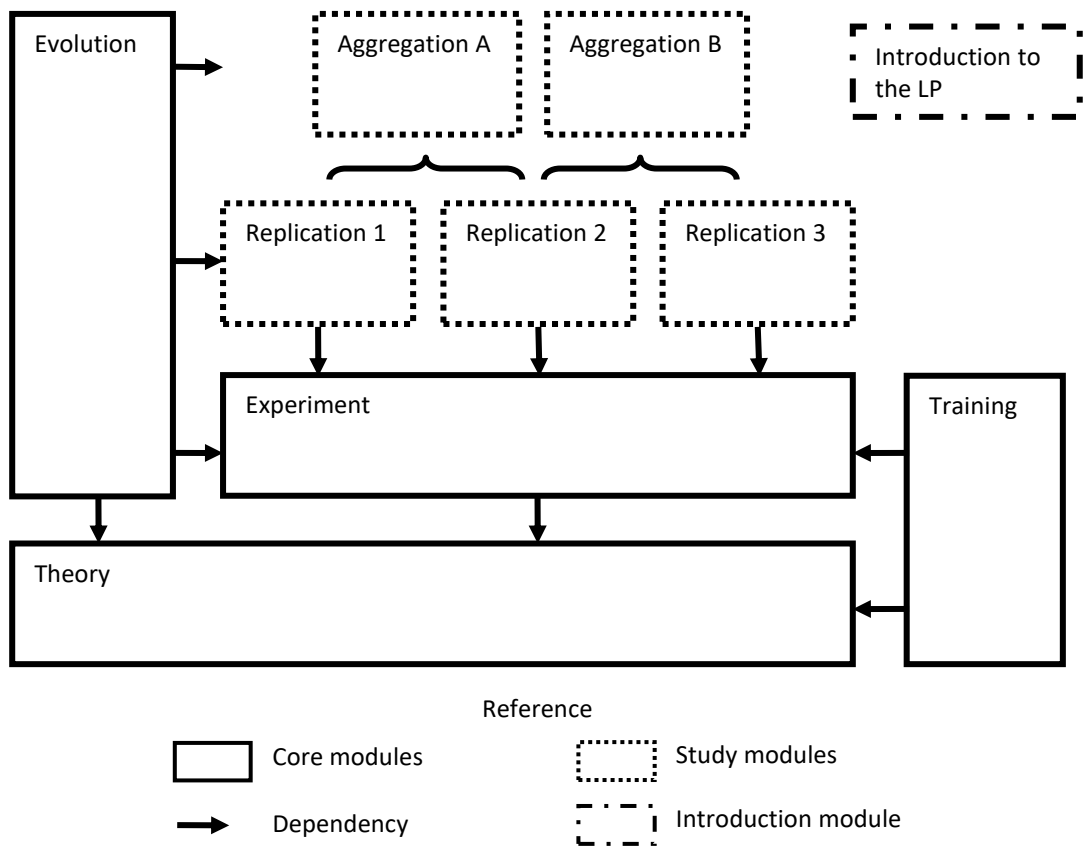

**Figure 1: Proposed LP modules.**

The **introduction to the LP** module contains information about the LP itself. This module is at a different level of abstraction to the other modules. Whereas the other modules refer to the experiment, this module refers to the actual LP. The Introduction module is designed to provide the LP user with information about the structure and use of the LP. The LP users may be researchers intending to replicate an experiment or researchers setting out to package an experiment for replication by others. The Introduction module denotes the aims of the instrument (LP), possible uses and how it can be enriched. It also includes the suggested communication instances and the particulars of the researchers who should be contacted in order to run a replication.

There are two other types of LP modules: **core modules** and **study modules**. The core modules are modules that contain basic information about the experiment and its underlying theory. While there is only one instance of these modules in the LP, there should be more than one version for the purposes of LP configuration management. The core modules must be modified when significant changes are made to the experiments, thus generating new versions. The study modules each match a replication or aggregation conducted as part of the family of experiments. The number of instances of the study modules in the LP is equal to the number of studies that have been conducted. The study modules are added to the structure of the LP with each replication or aggregation, without altering the previous study modules.

The **core modules** have a higher level of abstraction and contain the groundwork of an experiment. These modules include the theory, the instructions for replicating the experiment and training material. The module with the highest level of abstraction is the experiment **theory**. It establishes the theoretical framework for conducting the experiment and interpreting the results. All the other modules are directly or indirectly related to the theory module. The **experiment** module defines the experiment to be replicated. This module includes the instructions for the replicating experimenter, the operational material and the experimental tools. The **training** module contains all the materials used to train the experimental subjects (if applicable).

The **study modules** are added to the LP as the replications or aggregations of the experiment are conducted. Instances of these modules can be added for every replication or secondary study carried out with LP data. The **replication** modules capture information about each replication. The replicating experimenter is expected to add a module of this type describing his or her replication to the LP. Each module includes the description of the adapted design, the resulting data and notes on the replication experience. The **aggregation** modules are for recording the secondary analyses conducted on the replication data. These modules record anything from the inspection protocols, meta-analyses or other comparison techniques used and the findings of the secondary study.

The **evolution** module is a special case within the LP. There is only one instance of this module, and it is therefore an experiment module. However, a new section is aggregated to the evolution for each and every one of the replications and aggregations added to the LP. This module serves the purpose of establishing a relationship between the experiment versions and their replications, as well as of picturing how the LP evolves as a whole. For example, when several replications are conducted by different research groups, the exper-iment often tends to evolve and the experimental objective to change (in order to further generalize the results). This module should reflect the historical evolution of the experiment and the replications to which each version of the experiment is linked.

When it is necessary to modify any part of the experiment, the changes should be made within one and the same module whenever possible. However, there are often external dependencies. In order to make the LP easier to read, module dependencies are one-way only. The lower-level modules depend on the higher-level modules. For example, the replication modules depend on the experiment module, and this depends on the theory module. Generally, this avoids circular references and eases LP evolution.

## 5.2 LP template

Each module of the proposed LP structure is further divided into sections and subsections. These sections are described in detail in a LP template document. The LP template is a tool for the researcher who is responsible for instantiating an experiment. The LP template assigns a correlative number to and provides a content description for each section and subsection. Table 4 shows an excerpt of the LP template for the experiment module. It gives a sample content description for one subsection in each section of the module, omitting the other subsections for brevity. The LP template document with the content description for each section is included in the web appendix.

**Table 4: Excerpt of the LP template for the experiment module.**

| Section | Subsection | Content |
|---|---|---|
| **4.1 Planning** | *4.1.1 List of replication activities* | Each activity required to perform the replication must have a description, specifying the dependency and order of the activities for the replicating experimenter. |
| | *4.1.2 Estimated workload* | (…) |
| | *4.1.3 General schedule* | (…) |
| **4.2 Study conception** | *4.2.1 Objectives* | Description of the high-level attributes and goals examined by the experiment. |
| | *4.2.2 Hypotheses and sub studies* | (…) |
| | *4.2.3 Factors and response variables* | (…) |
| | *4.2.4 Contextual variables* | (…) |
| **4.3 Experimental design** | *4.3.1 Design alternatives* | List of design alternatives for the experiment. The alternatives can include experimental designs already used in earlier replications that are potentially valid bearing in mind the study conception. They should include a reference to the experimental theory used as a basis for the design. |
| | *4.3.2 Guidelines for selecting the experimental design* | (…) |
| | *4.3.3 Validation of the experimental design* | (…) |
| **4.4 Operation** | *4.4.1 Instructions for preparing material* | Steps necessary for preparing the material for the sessions. This may involve printing out forms or setting up a laboratory environment. They must specify what steps the replicating experimenter can take to improve the management of the material and assure the correct application of the experimental design. They should include a checklist of deliverables that may vary depending on the treatment. |
| | *4.4.2 Operating material* | (…) |
| | *4.4.3 Instructions for running sessions* | (…) |

| *4.5 Analysis* | *4.5.1 Data collection* | This section specifies the data collection method and any changes necessary to output the response variables. In some cases, several measurements may have to be taken to measure a response variable. All units and transformations must be specified. The final data are collected using standard templates with data fields for each factor and response variable. A spreadsheet template or similar should be included to make the replications easier to compare. |
| --- | --- | --- |
| | *4.5.2 Analysis methods* | (…) |
| | *4.5.3 Results interpretation* | (…) |

To use the LP template, we suggest that researchers start with simple instantiations of the core modules. The experiment module aims to capture the essential information needed to replicate an experiment. The other core modules —theory, training, evolution— should be used when needed to capture additional knowledge or when the experiment module contains a lot of information. These modules could be built after the initial effort to understand the baseline experiment. The training module is designed to include all the materials used in training. We have found that, in academic settings, experiments are usually conducted as part of a course, and it is common for educational materials (for both students and teachers) to add value, providing context for the replicating researcher.

## 5.3 Packaging checklist

The proposed LP structure is the groundwork for deploying other improvements to content and writing style. The incidents identified in the diagnosis stage show that many of the problems in the use of the LPs are due to information that is missing or not detailed enough for replication. The writing style is crucial for information comprehensibility and applicability. LPs should be written as *step-by-step instructions* focusing on activities that the replicating experimenter has to perform. On this ground, we drew up a number of documents detailing the LP structure and further specifying packaging criteria.

Apart from the LP template document, we provide a packaging checklist to be used by the researcher instantiating a LP for a specific experiment. The packaging checklist summarizes all the proposed improvements to the structure, content and writing style. The checklist is divided into four different key SE experiment packaging categories:

- Instructions for the replicating experimenter

- Operational material

- Experimental research process support

- Structural and usability improvements

Table 5 shows examples of checklist items for each category. The specific components of each improvement are added to the packaging checklist to assure that the researcher can easily check the instantiated LP. Some of the improvements are already addressed by the modularity of the LP template or in specific LP template sections. However, the packaging checklist could be used to check and improve the first draft of the LP. The packaging checklist document with the components of each proposed improvement is included in the web appendix.

**Table 5: Excerpt from the packaging checklist.**

| Category | Proposed improvement | Components | LP template sections |
|---|---|---|---|
| Instructions for replicating experimenter | Replication plan (RP) | • List of activities and dependencies<br>• Estimation of times and resources by activity<br>• Basic replication schedule | 4.1 Planning |
| | (…) | | |
| Operational material | Sessions with time limit (ST) | • Maximum session time for subjects<br>• Short time limit (sessions lasting two or three hours) | 4.4.3 Instructions for running sessions |
| | (…) | | |
| Experimental research process support | Replication report (RR) | • Replication template: identification, characterization, results and lessons learned<br>• Modules added to the LP for each replication | 6.n.1 Description of the replication |
| | (…) | | |
| Structural and usability improvements | Navigation and search (NS) | • Conventional structures (index, table of contents, sections)<br>• Hyperlinks<br>• External references management<br>• Search engine | (*) Transversal to the whole LP |
| | (…) | | |

The following discussion of the proposed improvements included in the different sections of the checklist serves to illustrate how to use the LP template and packaging checklist. For example, one of the identified experimental incidents was that experiment replications are very time-consuming and that the total workload is hard to estimate. On this ground, we suggest that the LP should contain a **replication plan**. This component is detailed in section *4.1 Planning* of the LP template.

Another example of an identified incident concerned experimental objects and session scheduling: replicating experimenters had trouble replicating the experiment within courses with time constraints. On this ground, we suggest that the operational material of the LP should include, whenever possible, other experimental objects (programs, models, test cases, etc.) to be used in **sessions with time limits** that can be adapted to the courses. This component is detailed in section *4.4.3 Instructions for running the sessions* in the LP template.

In a further example, the original DefDetect experiment LP contained no guidelines on how the replicating experimenter should report the replication. As a result, incidents such as very late replication reporting in wide-ranging formats were observed. This is an obstacle to the advancement of knowledge. On this ground, one of the improvements that we propose is a guideline for **replication reporting**. This component is detailed in section

*6.n.1 Description of the replication* (*n* will be the number of replication study) of the LP template.

Regarding the LP format, some replicating experimenters drew attention to the fact that a lot of the material is hard to use. In view of the state of the art with regard to documentation, we suggest a minor improvement whereby the document should include **navigation and search facilities**. This proposed improvement is implemented in the structure itself and not in a particular section of the LP template.

# 6  Generating LPs for experiments

We arranged the experimental materials that the group of researchers had put together according to the proposed LP. This process relies on access to the knowledge on each experiment, including tacit knowledge that may not be openly documented. Even though the diagnosis stage improved our direct knowledge of the experiment enormously, we had to examine the available materials more thoroughly and meet with the experimenters in order to instantiate a LP.

In the case of experiments with which the person who built the LP was unfamiliar, this turned out to be a very demanding process. This makes sense since the researcher preparing a LP is expected to and should have been involved in running the experiment so it should not be considered a threat to our proposal. Two LPs were instantiated in Iteration 4. For the experiment on team and personality influence in development (TeamPersInf experiment), the researcher responsible for instantiating the LP was unfamiliar with the experiment. For the experiment on unit testing techniques (UnitTest experiment), the researcher responsible for generating the LP was unfamiliar with the LP structure proposal but had been involved in the experiment.

To help researchers (other than the authors of the proposal) apply the LP instantiation to different experiments, we generated a LP **instantiation procedure**. The aim of the procedure is to improve instantiation by providing support for processing the available materials (that may or may not be part of a pre-existing LP), as well as eliciting any tacit knowledge that needs to be specified. Figure 2 outlines the LP instantiation procedure.

The main sources for instantiating a LP are existing documents related to the experiment and tacit knowledge. The instantiation procedure relies on two tools that are part of the LP proposal: a LP template and a packaging checklist. After a number of intermediate results (restructured documents, list of missing components and new components), the end product is a LP instantiated for a specific experiment.

The first activity of the instantiation procedure involves **gathering existing documents** and material about an experiment. The LP template is an essential tool in this activity, since it details every section that should be included in a complete LP. For a person less experienced in experimentation, the LP template provides support in two ways. First, the explanations included for each section of the LP template inform about experimentation concepts. Second, the section structure guides in the identification of material about the experiment that should be gathered if available.

The LP template is used in the **completeness check** activity to generate a list of missing components. Using this list as a reference, the **generation of new components** activity involves interacting with researchers who may have the required knowledge and producing new components for the LP. Like any knowledge gathering procedure, the activities do not necessarily occur sequentially. However, by first analysing existing documents and checking the completeness against the proposed LP template, it is easier to identify the tacit knowledge to be gathered.

The packaging checklist is an additional tool for verifying the outcome. In the **review of changes activity**, the restructured documents and the generated components are checked against the packaging checklist. Many of the concepts included in the packaging checklist are transversal to the LP and not related to a specific section. For example, the writing style and the document format apply to the whole LP. The goal of the review is to improve the generated LP, not only aggregating existing documents and filling missing sections. As a result, researchers may be required to redo previous activities of the LP instantiation procedure.

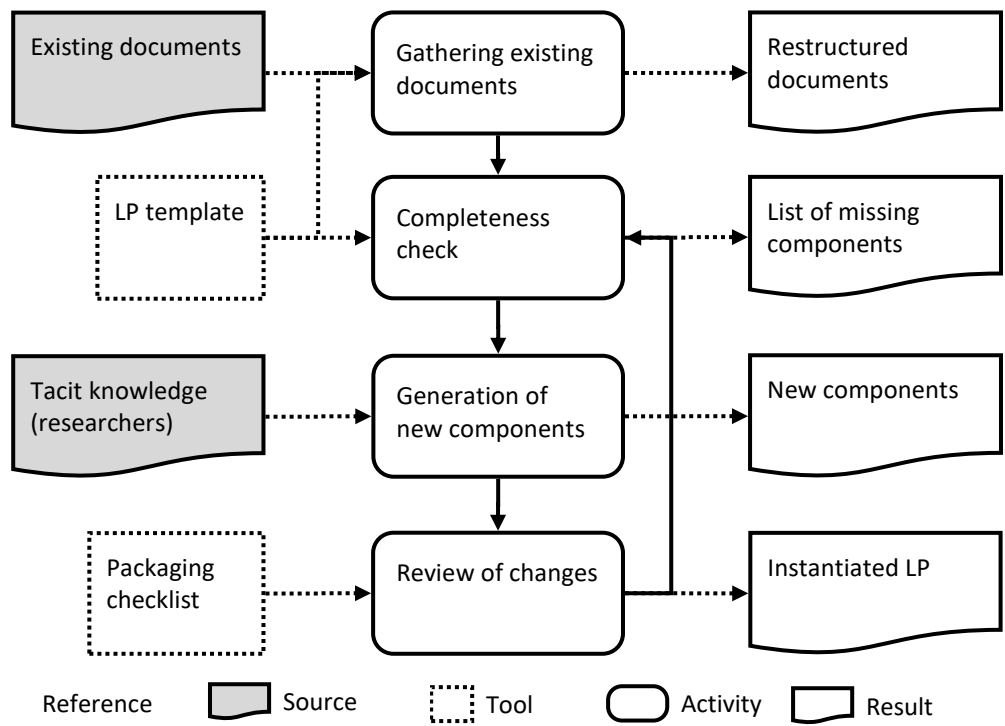

**Figure 2: LP instantiation procedure.**

The experiment used as a case study for identifying the experimental incidents (Def-Detect) was the first family of experiments for which the proposed LP was instantiated. We (the authors of this article) are very familiar with this experiment. However, a new LP was instantiated in order to adapt the materials to the proposed LP designed in response to incidents detected in several previous replications. The instantiated LP was used in replications by researchers not familiar with the experiment.

The second instantiated LP is a quasi-experiment on team and personality influence in development (TeamPersInf experiment) [50]. Random assignment is not possible in quasi-experiments. This experiment has several features that make it interesting for validating the proposed LP. Firstly, it is a complex quasi-experimental study, which includes a great many measured and analysed variables, measuring instruments, and instructions for analysing and interpreting results. The theory behind the study is backed by both psychology and SE. On the other hand, this is an advanced piece of research which has been internally replicated on a number of occasions, and has been reported in several published articles and a PhD thesis. It can therefore be considered as a mature piece of research providing all the knowledge required to elaborate a LP designed to externally replicate the study [57,58].

The third LP instantiation was conducted as part of an experiment on software unit testing techniques (UnitTest experiment) [59]. While this experiment studies testing techniques like the experiment that we used to identify incidents in replications, it is part of a different family of experiments conducted by another research group. The instantiation of this experiment has two goals: validate the proposed LP and put together a LP for use in future replications. From the viewpoint of the experiment, we took advantage of the validation to organize the information on an experiment and make it available for use in coming replications. The results of this instantiation with regard to both objectives were reported in [60]. In this case, the LP was instantiated by a researcher that did not author the LP proposal but was involved in running the experiment.

For the LP generation, we used a digital document that contains all the instantiated LP modules. On top of this, the operational material and the datasets from the replications are saved in a directory structure. One possible improvement in this respect would be to use web standards and a collaborative software infrastructure associated with the LP. At this stage of the research, we focused on the issue of structure and content irrespective of the implementing technology and used just a digital document and materials directories. Posting information about an experiment on the web makes it accessible to a wider audience. Unless it has the right structure and content for replication, however, its being on the web will not improve matters any, that is, as we have repeatedly found with published LPs like [16], the web does not solve the problem of which information a LP should contain.

# 7 Evaluation of the LP proposal

In the evaluation stage, we validated our LP proposal. On one hand, we checked that the LPs instantiated according to the proposal were useful for the research groups involved. On the other hand, we collected evidence about how the proposed LP supports conducting effective replications of experiments and reduces the number of incidents in the replications. Our evaluation observed the use of the proposed LP structure in different experiments and in different replication scenarios. The collected evidence shows the initial validity of the proposed LP and how it can be iteratively improved.

Our proposal was evaluated by dividing the overall objective of an LP into more specific aspects. We used two complementary approaches to evaluate the proposal: LP instantiation for different experiments and use of the instantiated LPs in replications. On one hand, by instantiating the LP proposal, we check whether the proposal is capable of generating a LP for different families of experiments, that is, that the LP structure that we propose is **feasible** and **flexible.** On the other hand, we evaluate replications of experiments using LPs instantiated according to our proposal. By using LPs in replications, we can analyse how much support they provide for the experimental process activities, as well as how the replicating experimenter rates the quality of the document, that is, that the LP that we propose is **complete** and **usable**. These replications were evaluated according to the same procedure as in the diagnosis stage in order to compare the impact of using our proposal in experiment replications. By thoroughly observing the use of the LP in replications (including any experimental incidents), we can evaluate the **efficacy** and **efficiency** of our proposal, as well as its impact on the **satisfaction** of the replicating experimenter.

## 7.1 Evaluation of our LP proposal

The specific aim of the first evaluation activity is to check the **feasibility** of our LP proposal. We understand that the proposal is feasible if it can be instantiated as a LP for a specific experiment with a workload and outcome that is acceptable for experimenters. In the first iteration, we instantiated the DefDetect experiment that our research group is working on and wants to improve.

The **feasibility** criterion is divided into three more specific attributes: instantiation, instantiation workload and generated LP acceptance. We measured the workload attribute in person-hours. For the instantiation and acceptance attributes, we used a binary scale to simplify the evaluation process: either the generated LP was instantiated and accepted by the experimenters or not. Other aspects of the generated LP, as the document quality and ease of use are evaluated by other attributes. The result of this evaluation activity was the successful generation of a LP instance for the DefDetect experiment in return for a reasonable amount of effort on the part of the experimenters. On this ground, we consider our LP proposal to be workable.

We then instantiated our LP proposal for the TeamPersInf and UnitTest experiments. These instances have the potential of corroborating the feasibility evaluation of the proposal. However, the key goal of instantiating these LPs is to check that the proposal is **flexible** enough to adapt to different experiments. Many things may vary from one experiment to another and from one context to another: the characteristics of the experimental study, the knowledge available for instantiation and the context in which the experiment is run, among others. In order to evaluate flexibility, we thoroughly examine the differences between the generated LPs to check whether they account for the specified characteristics of each family of experiments. In particular, we looked at whether sections had to be added or removed in order to instantiate a LP for each experiment.

The flexibility of the proposed LP is not an absolute term. We found that a LP instantiating our proposal can be generated for three families of experiments, whence we inferred that the proposal may be applicable in other cases. As LPs are generated (by instantiating our proposal) for more experiments, the flexibility of our proposal is further corroborated. If the LP is instantiated for other families of experiments, our proposal may require adjustment. Each experiment is likely to have particular characteristics and require a specific treatment not accounted for in the original proposal. Such changes are unavoidable, as a high-level solution cannot be expected to account for each and every case. Any such changes that occur will have to be documented and analysed. Whenever appropriate, they may be integrated into the basic LP proposal.

The TeamPersInf and UnitTest experiments differ from DefDetect experiment in different respects. The TeamPersInf experiment deals with another process area within SE. The study method is a quasi-experiment, which is quite far removed from the original controlled experiment. As the UnitTest experiment this is again a controlled experiment on software testing techniques, the differences have less to do with the topic of study or the experimental design than with the person conducting the instantiation and the knowledge available about the experiment.

In both instantiations, we observed that the instantiation effort is within an acceptable range (from 75 to 85 hours). This considers the amount of effort required to adapt the LP structure and generate new documentation for each experiment. This observation validates the feasibility issue already confirmed by the first instantiation. We also found that there is not a big difference in the instantiation effort for a person unfamiliar with the proposed LP.

## 7.2 Evaluation of the use of the LP in replications of experiments

The aspects evaluated in this activity focus on the use of the LP during replication. In particular, we evaluate the **completeness** and **usability** of the documentation based on our proposed LP. We examine whether it supports all the activities of the experimental process and how usable it is from the viewpoint of the researcher who uses it to replicate an experiment. On the other hand, we evaluate the **efficacy** and **efficiency** of the replication, and the subjective experimenter **satisfaction**. It is hard to evaluate these aspects through a simple static analysis of the instantiated LPs. Ideally, the use of the LP should be observed within the framework of an experiment replication.

In order to evaluate the replications, we applied the same empirical evaluation procedure to the replications as we used in the diagnosis stage of this research. By strictly adhering to the same evaluation procedure, the result will be comparable to the replications evaluated in the previous iterations. Specifically, we can observe whether the LP prevents or mitigates the experimental incidents identified in replications where our LP was not used. This does not mean that we have found the ultimate solution to the LP problem, but it does at least raise the confidence that we have in our proposal improving upon the current quality of the information used to replicate experiments.

The evaluated replications used LPs instantiated according to our proposal for the DefDetect experiment, the same experiment used in the replications of the diagnosis stage of Iterations 1 and 2. The exact objective of using the same experiment is to improve the LP for the research group. On the other hand, by using an improved LP in later replications of the same experiment, we were able to compare the impact of LP use directly.

The evaluation included two replications, each of which was performed in a different context and by a different replicating experimenter. The replicating experimenters involved in each replication are unfamiliar with the details of the experiment. This is a necessary condition of the evaluation, as, if they were acquainted with the experiment, we would not be able to evaluate the impact of the LP as a knowledge transmission instrument. In both cases, we made contact with the participating researchers, thanks to which we were able to evaluate the replication at close quarters, albeit without intervening in its execution. Running two replications will increase the validity and reliability of the results.

The experimenter using the LP in each replication rated document usability directly using a five-value Likert scale, where 5 is best and 1 is worst. On this ground, although similar LPs were used, the rating varies considerably across replications. However, we can get an idea of document usability as perceived by the participating researcher by taking the mean rating scores.

Table 6 shows the values for the usability attributes and components of all the evaluated replications. The columns shaded grey refer to the replications using LPs structured according to this proposal. For the purposes of comparing the numerical usability values, the table includes two columns listing the means before (PRE) and after (POST) the application of the proposed LP. The highest scores for all the usability attributes and components were recorded when the LP proposal was used. This applies to differing extents depending on the different attributes. We can, however, conclude that the usability of the document improves if the proposed LP is used.

**Table 6: Summary of LP usability evaluation.**

| Attribute | Component | UPM 2000 | UPM 2004 | UdS 2005 | ORT 2005 | UPV 2006 | ORT 2009 | UPM 2011 | Mean PRE | Mean POST |
|---|---|---|---|---|---|---|---|---|---|---|
| *Ease of application* | Task orientation | 2 | 1 | 5 | 2 | 4 | 5 | 4 | 2.8 | 4.5 |
| | Accuracy | 4 | 3 | 3 | 4 | 4 | 5 | 4 | 3.6 | 4.5 |
| | Completeness | 2 | 1 | 3 | 2 | 3 | 5 | 4 | 2.2 | 4.5 |
| *Ease of understanding* | Clarity | 3 | 1 | 5 | 3 | 4 | 4 | 5 | 3.2 | 4.5 |
| | Concretion | 3 | 1 | 4 | 2 | 5 | 4 | 5 | 3 | 4.5 |
| | Style | 3 | 1 | 5 | 4 | 4 | 4 | 4 | 3.4 | 4 |
| *Ease of search* | Organization | 3 | 1 | 5 | 3 | 5 | 4 | 5 | 3.4 | 4.5 |
| | Retrievability | 3 | 3 | 5 | 2 | 4 | 4 | 3 | 3.4 | 3.5 |
| | Visual effectiveness | 1.5 | 1 | 3.5 | 2 | 2 | 4 | 4 | 2 | 4 |
| Values assigned by the experimenter using the LP (5-point Likert scale where 1=worst, 5=best) | | | | | | | | | | |
| Columns shaded grey refer to the replications using LPs structured according to this proposal | | | | | | | | | | |

Table 7 shows the values for the completeness attributes and components of all the evaluated replications. The columns shaded grey refer to the replications using LPs structured according to this proposal. The comparative study of completeness illustrates the evolution of the LP. The replications using LPs structured according to the proposal provide more coverage of experimental process activities. The scope and level of instruction of these LPs is better. Albeit to a lesser extent, we found that experimental designs are more easily adaptable using the LP proposal. Using the structured LPs, a replication report can be drawn up faster and then integrated into the LP itself. We can conclude that the LP structured according to the proposal is more thorough than the non-structured variants.

**Table 7: Summary of LP completeness evaluation.**

| Attribute / *Component* | UPM 2000 | UPM 2004 | UdS 2005 | ORT 2005 | UPV 2006 | ORT 2009 | UPM 2011 |
|---|---|---|---|---|---|---|---|
| Scope<br>Values: Training, Design, Operational, Complete | Operational | Training Design Operational | Training Design Operational | Desiin Operational | Design Operational | Complete | Complete |
| Level of instruction<br>Values: Materials only, Basic, Detailed, Detailed – Grounded | Materials only | Basic | Detailed | Basic | Detailed | Detailed - Grounded | Detailed - Grounded |
| Adaptability<br>Values: Unchangeable, Partially adaptable, Adaptable | Unchangeable | Unchangeable | Partially adapt. | Unchangeable | Partially adapt. | Adapt. | Partial adapt. |
| Scalability / *Replication reported*<br>Values: Yes, No, Late | Late | Late | No | Late | No | Yes | Yes |
| Scalability / *Aggregated results*<br>*Values: Yes, No, Partial* | Partial | Yes | No | No | Partial | Partial | Partial |
| Version control<br>Values: No, Current version only, Log, Retrievable versions. | No | Current version only + log | Current version only | Current version only | Current version only | Current version only | Retrievable versions |
| More than one element could be assigned for the Scope attribute (Complete includes all elements).<br>For the other attributes, the possible values are ordered worst to best. | | | | | | | |
| Columns shaded grey refer to the replications using LPs structured according to this proposal | | | | | | | |

The main objective of an LP is to answer replicating researchers' questions in order to reproduce the intended environment for the experiment. An improved LP should mitigate experimental incidents, lowering the mean error severity. The efficacy of the LP as a knowledge transfer instrument was evaluated by means of three qualitative attributes (as in the diagnosis stage): question answering, environment reproduction and mean severity of experimental errors. To compare each replication, we used a three-level scale to measure each attribute. Error severity is regarded as high if an experimental incident prevents the integration of the experimental results into the experimental body of knowledge of the respective research project. Error severity is medium when the replication has a sizeable number of incidents, and additional effort is required for results integration. A low mean error severity occurs when the incidents do not much alter the course of the replication and do not affect the results. We used the same replication evaluation instruments and procedure to determine these qualitative attributes in both the diagnosis and the evaluation stages (semi-structured interviews with experimenters, document analysis and additional questions).

Table 8 shows a summary rating these attributes across seven evaluated replications. The two replications that use the LP instantiated according to our proposal are highlighted in grey. The two replications conducted with the LP based on our proposal yield better results than the previous replications. In practical terms, this means that we identified fewer incidents overall and the severity of the incidents was lower.

**Table 8: Summary of LP efficacy evaluation.**

| Iteration | Iteration 1 | Iteration 2 | | | | Iteration 3 | |
|---|---|---|---|---|---|---|---|
| Research stage | Diagnosis | Diagnosis | | | | Evaluation | |
| LP version | Without LP | Previous version | | | | Our proposal | |
| | UPM 2000 | UPM 2004 | UdS 2005 | ORT 2005 | UPV 2006 | ORT 2009 | UPM 2011 |
| Question answering | Low | Complete | Medium | Low | Medium | High | High |
| Environment reproduction | N/A | Complete | Low | Low | Medium | Complete | Complete |
| Mean error severity | Medium | Slight | Serious | Medium | Medium | Slight | Slight |
| Columns shaded grey refer to the replications using LPs structured according to this proposal. | | | | | | | |

During the replications, we also found that the usability and completeness of the document used as the LP was perceived to be more positive than in the replications that do not use the proposal. These issues were evaluated taking into account the opinion of the researcher involved in the replication (none of whom were involved in putting together the proposed LP) and the ultimate analysis of the document content.

The efficiency was measured by asking researchers replicating the experiments using the improved LP to log her effort in person-hours. Reliable estimations of the effort needed for replicating the experiment before the LP proposal were not available. However, we tried to reconstruct the replication process in the interviews with the replicating experimenters. Although we are not able to provide a rigorous evaluation of the efficiency of the proposed LP, we did observe that the effort was lower to the estimate effort for previous replications.

Each evaluation creates a detailed analysis of incidents and LP use. These experiments and replications checked that our proposal meets the following conditions:

1. The LP can be instantiated according to our proposal for different SE experiments (feasibility and flexibility).

2. The quality of the instantiated LPs is satisfactory (completeness and usability).

3. The use of the LP improves the experimental process (efficacy, efficiency and replicator satisfaction).

As limitations of this proposal, we should note that it only provides partial support for aggregation studies. It is possible to use the template to include secondary studies, but this has not been validated in either of the two replication studies. The version control proposal (including multiple versions of the experiment in the same LP) was used in one of the replications. However, we found that these multiple retrievable versions were not used to plan the actual replication. Experiment change management is a much more complex issue that we expected. Therefore, another separate research project was initiated to further explore this issue.

We admit that less experienced researchers might find it challenging to apply the LP structure proposal. However, the researchers responsible for the evaluated replications were graduate students without much experimental experience. Overall, they successfully instantiated the LP and replicated the experiment without direct intervention.

## 8 Generalization of our LP proposal

Experiment replication is a demanding process. On top of this, experimenters and information on an experiment are not easily accessible when conducting research like this. Three experiments were considered throughout this research. We looked for experiments that differed significantly with respect to the procedures and materials used. Even so, not many changes had to be made to the LP. Taking into account just how limited access to experiments is, we consider that the above three experiments constitute a big enough sample for us to be able to claim that our LP proposal is sufficiently generalizable across experiments.

Another component of the replication, apart from the experiment itself, is the replicating researcher. This is a key component of our research, as the researcher is precisely the knowledge transfer target. Subjectivity comes into play when each researcher interacts with the LP. To address this concern, we engaged seven different replicating experimenters in the replications analysed in our research. One necessary premise for evaluating the impact of the LP is that researchers should be unfamiliar with the experiment at the time of conducting the replication in question. This means that we can study the most troublesome (but quite common) instance of information transfer in a replication: when the replicator is completely unfamiliar with the experiment. Besides, the seven researchers have differing levels of experience in experimentation: some are experts, whereas this is their first contact with a SE experiment for others. In all cases, we gathered information about other mechanisms of communication apart from the LP that were used in each replication. Although we cannot control all human factors and means of communication having an impact on a replication, we believe that we have mitigated this threat by accounting for a diverse set of replicating experimenters and replications and, as far as possible, taking note of the knowledge transfer instruments. Therefore, we take the view that, having studied seven different replicating experimenters in seven replications, the generalizability of our proposal with respect to experimenters is acceptable.

The findings with respect to the LP structure and content are limited by the number and diversity of the experiments packaged using our proposal, as well as by the researchers participating in the replications studied. Across the research iterations, three different experiments were packaged and seven replications conducted by seven different researchers at five different sites were evaluated. Thus, we are confident that our LP proposal is generalizable enough to be able to be applied to other SE experiments.

Regarding the cultural environment of this research project, all the evaluated replications were conducted in Spanish. The LP proposal and its instantiation for specific experiments were also written in Spanish. In view of these constraints, the transferability of the proposed solution to other environments requires further consideration. We do not think that the LP proposal has any cultural or language-specific terms. However, the translation of the LP template and packaging checklist should be evaluated in practice by means of a pilot replication study.

One validity threat to our proposal is our connection with the authors of the experiments and evaluated researchers. Two of the authors of this paper were responsible for two replications (Vegas in UPM 2000 and Solari in ORT 2005). The person that collected and analysed the data for evaluating all the replications (Solari) was also a member of the team that put together the LP proposal. There is, therefore, a risk of self-confirmation in any case where a researcher tests his or her own method or solution. To mitigate this threat, we defined a replication evaluation procedure beforehand. This procedure specifies activities, intermediate products and coded value levels. The results can be traced back to and reproduced from the defined information sources. For the open coding, we linked all the codes to literal citations from interviews and documents applied in the replication. The results of each replication evaluation were validated by the principal investigator responsible for the replication or the experiment. At the validation stage of the proposed LP, we took care not to intervene in the replication so as not to influence knowledge transfer in any way. However, this threat cannot be completely eradicated because the replicating experimenter is aware that the replication and particularly the LP are being evaluated.

We do not claim that the resulting proposal is universally applicable or succeeds in fully supporting replication. As opposed to pure scientific research, the aim of technological research is to find solutions that improve upon the current situation. The evaluation that we conducted shows that the proposal is acceptable and to some extent reduces the amount of effort currently required to replicate experiments in SE. We do not claim that this proposal is the only or even the best possible LP structure. It is a structure that has been satisfactorily validated and can thus be recommended for use in practice.

The further validation and evolution of this proposal requires instantiations of experiments from different SE domains and more replications using structured LPs. We would like to call on collaborators to add more experiments. All the elements of the proposal are publicly available from our website[14]. We would be happy to provide support for other researchers using the LP template and packaging checklist to instantiate their own experiments. We are also willing to instantiate a LP for experimenters who send in experiment materials. Experimenters would only be required to be available for tacit knowledge elicitation. We would like to receive any feedback on the instantiation process and, if possible, evaluate the replications using the structured LP. Each and every experiment and replication context will improve the proposal and foster collaboration among experimenters to replicate experiments.

# 9  Further evolution of the LP proposal

This version of the LP proposal is implemented in a set of files (Introduction, LP template document, Packaging checklist document). We chose this format for simplicity: the single file documents can be downloaded and edited using a popular word processor without requiring specific knowledge. However, we think that future versions of the LP template should be based on Web standards. The hypertextual support provided by Web standards would be better suited to the modular LP structure. A standardized format would improve the general portability of the LP. However, this change should be evaluated carefully against the availability and ease of use of editing tools.

---

[14] http://www.grise.upm.es/LaboratoryPackage

Another possible way of evolving the LP proposal would be its implementation within an online collaboration platform. The platform would enable the creation and publication of LPs for different experiments that could be easily transferred and used. It would also provide communication among researchers in parallel to LP use. All the participants in an experimentation project would be able to view questions and answers. The ultimate aim of this proposal is to foster replications that can be integrated into a SE body of knowledge. An online collaboration platform would speed up that process. On the other hand, such a platform would raise user authentication, security and other issues that can be hard to implement.

A LP is used in scenarios where there are different levels of communication among researchers. We have found that it is harder to understand the LP and conduct the replication if there is no communication between the experimenters running the baseline experiment and the replication. The results of a replication where there has been no communication among experimenters are harder to interpret and integrate into the broader experimental body of knowledge of the respective line of research. We believe that this LP proposal could yield better results if it is used in combination with communication mechanisms while performing the replication. We have found that an experimental design validation meeting and a results interpretation meeting at least are necessary to produce useful replications. The LP proposal could be further developed to complement the inter-researcher communication mechanisms. To do this, we suggest that the communications among researchers should be included in the LP history and the useful experiences should be assessed in order to evolve the communication section.

We decided to focus our LP proposal on controlled experiments. Controlled experiments are a common thorough approach within the experimental scientific method. Controlled experimentation requires the precise definition of techniques, measurement and context. In some cases, replications depend on the researcher conducting the training, assignment, monitoring or data analysis according to predefined criteria. During the evaluation phase, we instantiated the LP proposal for two controlled experiments and one quasi-experiment. We found that the proposal is flexible with respect to the experimental approach. However, other empirical studies may require different approaches. The LP would have to be instantiated for case studies and other observational studies with different degrees of intervention in order to further evolve the proposal. We believe that most of the modules in the LP template and the packaging checklist would be useful for a wide range of scenarios, but the proposal would have to be evolved based upon a thorough analysis.

# 10 Conclusion

Experimentation is a key strategy for maturing knowledge in any area of science and technology. SE needs experimentation in order to provide reliable evidence about techniques, methods and tools. However, an individual experiment does not provide unques-tionable and irrefutable results. Experimental research is an incremental process at the heart of which is the replication of experiments. Experiment replication cannot be done without the transfer of knowledge among researchers, and this is one of the major challenges now facing empirical research in SE.

If replications are to be less costly and complex, special-purpose instruments are required to carry out experimentation in SE. LPs are one of the principal instruments for supporting knowledge transfer and running replications. We have put together and evaluated a LP proposal for SE experiments. The purpose of this research is to demonstrate that the replications run using the experiment LPs instantiated according to our proposal are better than replications using unstructured or ad hoc LPs which are what our community tends to use now.

We applied a set of defined stages iteratively to generate and evaluate our proposed LP. Thanks to this approach, we were able generate LPs for different experiments, as well as a LP structure and content proposal for SE experiments. We have generated LPs for different experiments and evaluated replications using the instantiated LPs.

Throughout this research, we evaluated seven replications. These replications accounted for three different families of experiments and five replication sites. In all cases, the replications were conducted by researchers who were unfamiliar with the replicated experiment. In each case, we evaluated different aspects of our proposed LP. Our proposal proved to be good enough to iteratively improve these three families of experiments.